# Interactive Rainbow Score:
# A Visual-centered Multimodal Flute Tutoring System

Daniel Chin    Yian Zhang    Tianyu Zhang    Jake Zhao    Gus G. Xia

Music X Lab

New York University, Shanghai

{ daniel.chin, yian.zhang, tz1201, j.zhao, gxia } @nyu.edu

## ABSTRACT
Learning to play an instrument is intrinsically multimodal, and we have seen a trend of applying visual and haptic feedback in music games and computer-aided music tutoring systems. However, most current systems are still designed to master *individual* pieces of music; it is unclear how well the learned skills can be *generalized* to new pieces. We aim to explore this question. In this study, we contribute Interactive Rainbow Score, an interactive visual system to boost the learning of *sight-playing*, the general musical skill to read music and map the visual representations to performance motions. The key design of Interactive Rainbow Score is to associate pitches (and the corresponding motions) with colored notation and further strengthen such association via real-time interactions. Quantitative results show that the interactive feature on average increases the learning efficiency by 31.1%. Further analysis indicates that it is critical to apply the interaction in the *early period* of learning.

## 1. INTRODUCTION
### 1.1 Background of Multimodal Music Learning

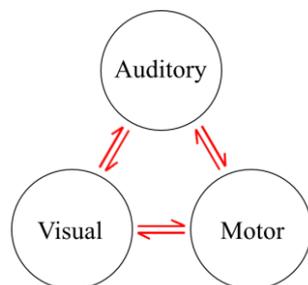

**Figure 1. A cognitive model of multimodal music learning.**

Learning to play an instrument is intrinsically multimodal [14]. It demands the human learner to incorporate multiple modalities, including visual (e.g. to see and read the musical notes), motor (e.g. finger movements on a piano), and auditory (e.g. to listen to the music one is playing). These modalities interact with one another as one plays music. Figure 1 shows the three modalities and their interactions.

There are usually two types of music learning skills. **Type I**: to learn *individual* pieces. This means to memorize some of the three representations of a piece (the music notation, the finger movements, and the sound) **Type II**: to develop *general musicality*. This means to learn the mappings (indicated by the red arrows) among the representations.

The key difference between these two types of skills is that the latter can be easily generalized to learn new pieces of music, while the former cannot. For example, in order to sing a song, one has to at least memorize the auditory representation, but such a memory would not help with learning another piece unless they two share a long subsequence of notes.

### 1.2 Interactive Multimodal Music Systems
There exist many music games and computer-aided music tutoring systems using multimodal feedback. For example, Guitar Hero [5], Rock Bands [1], and Taiko no Tatsujin [12] apply real-time visual feedback to indicate the timing of rhythmic performances. More recently, we see haptic instruments [14, 15] and gloves [8, 11, 15] being used in flute and piano tutoring. However, most systems are designed to help practice Type I skills; it is unclear how much of musicality, the Type II skills, can be gained from interacting with multimodal interfaces.

We aim to explore the question above. In this study, we focus on *sight-playing*, the general musicality to read music while mapping the visual representations to performance motions. To improve the learning efficiency, we contribute Interactive Rainbow Score, an interactive and intelligent visual interface inspired by chromagram [2] and piano roll [13] representations. This current interface is customized for *elementary* flute performance and has three key design features. First, pitch is associated with both color and note height for easy and intuitive score reading. Second, real-time visual feedback is displayed on the interface to further strengthen the visual-motor association. Third, several learning modes (with different levels of guidance) are created, and learners can choose which mode to use based on their learning progress. The user study shows that adding interactive features increases the learning efficiency by 31.1%.

In the next section, we describe the interface design in detail. We present the learning strategy in Section 3, the user study in Section 4, discussion in Section 5, and finally come to the conclusion in Section 6.

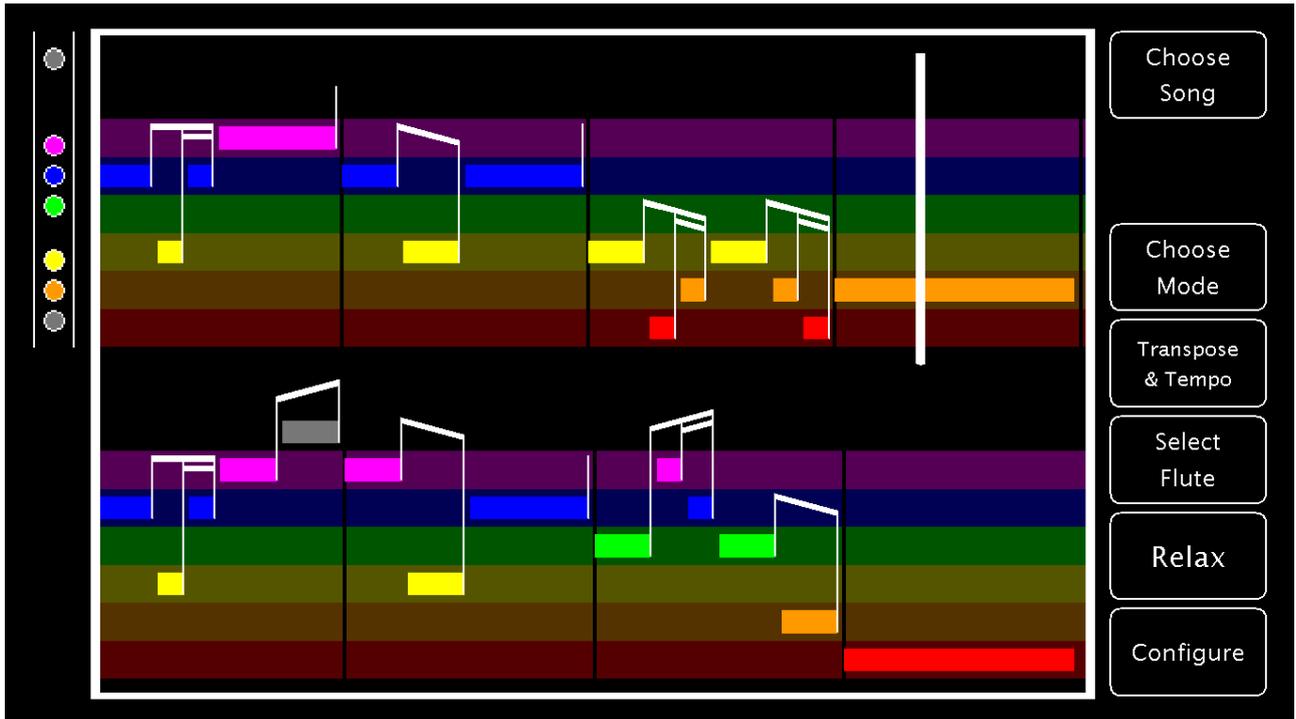

**Figure 2. An illustration of the Rainbow Score visual interface.**

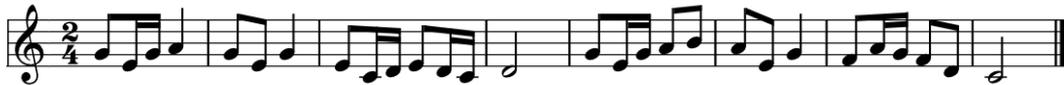

**Figure 3. The corresponding modern staff notation.**

## 2. INTERFACE DESIGN

We first present the Rainbow Score interface in Section 2.1 and then discuss the multimodal interaction feature in Section 2.2.

### 2.1 Rainbow Score

The visual interface is shown in Figure 2, in which an example of Rainbow Score notation is shown on the main panel laid in the middle, a flute icon is drawn on the left, and the control panel is displayed on the right. Each page is split into two rainbow bands for better displays. Each band is analogous to a staff and contains 4 chunks, each representing a *measure*. For example, the piece shown in Figure 2 contains 8 measures in total. (For ease of interpretation, the corresponding modern staff notation is shown in Figure 3.)

#### 2.1.1 Rhythm

Each note is represented by a colored rectangle, and the duration of a note is proportional to its width. The rhythm is also visualized using the stem-beam notation borrowed from modern staff notation. For elementary flute players, this design prepares them to later learn the modern staff notation.

#### 2.1.2 Pitch

As the current design focuses on elementary-level sight-playing, we restrict the pitch range to a diatonic scale in C major. To be specific, the seven pitches (C, D, E, F, G, A, and B) are laid out in 7 rows (from low to high), each assigned a unique color (red, orange, yellow, green, blue, purple, and grey) [7]. Thus, we see a trinity of pitch, color, and height, in which any one determines the other two.

Such trinity is further associated with performance *finger positions* via the flute icon, which contains seven holes: one blowing hole on the top and six finger holes below. The blowing hole is always grey, and each finger hole has two states: 1) *released*, if it is grey, and 2) *covered*, if the color matches the corresponding row. For example, in Figure 2, the playhead (vertical white bar) in the 4[th] measure indicates that the note being played is D, an orange note on the second lowest row. Such color and height information is further reflected on the flute icon — only the lowest hole is grey, which means that to play D on the flute, the player should cover all holes except the lowest one. Notice that B is triggered by releasing all finger holes and the top row has transparent background. Hence the Rainbow Score can be perceived as 6 rows for note C - A with an extra space above for note B.

In summary, the Rainbow Score notation inherits the benefits from both the abstractness of sheet music and the directness of finger notations such as guitar tabs. This advantage is the most obvious when compared to some commercial interactive pianos with a vertical piano roll falling downwards and keys lighting up for the player to press [10]. Such piano tutoring system has been criticized as musical whac-a-mole, in which the player remains passive and mechanical. Rainbow Score, on the other hand, requires the player to develop an abstract analogy between the visual representation and the motor movements. We believe this multimodal analogy is essential to sight-playing.

### 2.2 Multimodal Interactivity

To achieve better learning effect, we add interactivity to Rainbow Score, and name it Interactive Rainbow Score.

#### 2.2.1 Real-time Visual Feedback

On the visual interface, we use white masks to indicate player performance in real time. The white masks are placed beneath the colored notes and above the rainbow canvas. As a result, a correctly played note will be outlined with white borders, while a mistake will yield a jarring white mask, accompanied with arrows pointing towards the correct pitch. For example, the Interactive Rainbow Score in Figure 4 shows that the first 5 notes are played correctly but the 6th note E is incorrectly played as a D.

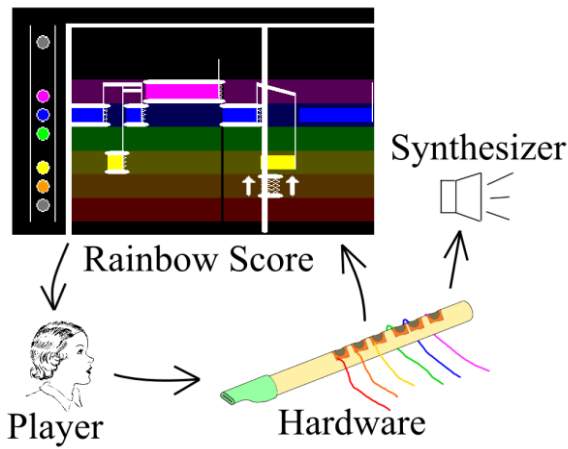

**Figure 4. An illustration of the overall multimodal interaction.**

### 2.2.2 Offline Mistake Review
When the player finishes a song, the entire performance history is displayed on the screen. The tutoring software then offers an option for the player to review the mistakes and toggle between the played version and the ground truth.

### 2.2.3 Hardware and Synthesizer
We adopt the hardware design in [14] and craft an electrical flute that reads the player's finger position using capacitive sensors and sends the real-time performance data to both the visual interface and the sound synthesizer. Similar to [14], the breath velocity is not measured, and the sound synthesizer assumes a constant breath velocity that restrains the performance in a single octave.

Above all, our system adds interactive visual feedback for sight-playing training on top of auditory feedback in the traditional setup. It is also different from most music games such as Guitar Hero [5] and Rock Band [1] in that most audio-visual effect of Interactive Rainbow Score is determined by human performance rather than pre-programmed. In essence, the multimodal feedback helps the player learn the mapping between visual notation and performance motion, and that mapping is still valid even when the feedback system is turned off.

## 3. LEARNING MODES
The Interactive Rainbow Score interface enables four learning modes, which are summarized in Table 1. Here, we see two perspectives: 1) static vs. interactive, and 2) system leads vs. performer leads. Among the four learning modes, A & B fully take advantage of the interactive visual feature, while C & D are considered the baseline.

**Table 1. A summary of the 4 learning modes**

|  | Interactive | Static |
|---|---|---|
| System leads progression | Mode A | Mode C |
| Performer leads progression | Mode B | Mode D |

### 3.1 Mode A: Frame-wise Feedback
In this learning mode, the playhead moves at a constant speed (which the player can set beforehand on the control panel) accompanied with a metronome played in the background. The player is supposed to follow the playhead in real time by performing the indicated notes. In the meanwhile, real-time visual feedback (introduced in Section 2.2.1) is displayed on the interface frame by frame, showing the actual performance so that the player can supervise the precise performance timing.

### 3.2 Mode B: Note-wise Feedback
In this learning mode, the playhead waits for correct performance. If a note is played correctly, it will be outlined with white borders and the playhead jumps to the next note. Otherwise, the playhead does not move and a white mask representing the learner input appears, alongside with a pair of arrows indicating the direction for the player to correct the note. (Again, see the Rainbow Score in Figure 4)

### 3.3 Mode C: Playhead Follower
This mode mimics some mainstream music games such as Taiko no Tatsujin [12], in which the player is also advised to keep up the performance with the playhead (same as Mode A) but there is *no* visual feedback.

### 3.4 Mode D: Free Practice
In this learning mode, the Rainbow Score remains static, and there is no playhead. The player is free to play any section of the piece at any speed. This mode mimics traditional music learning.

## 4. EXPERIMENT
To validate the effectiveness of the interactive visual feature on assisting sight-playing learning, we conducted a user study that compares the interactive learning modes A & B with the static baseline, modes C & D.

### 4.1 General Curriculum Design
It is important to remark that sight-playing is a general musical skill gradually gained through practicing different pieces of music. Therefore, we need more training pieces than related studies that aim to master individual pieces [14, 15]. To this end, we design a curriculum which consists of 16 short pieces, all modified based on folk songs.

During the training, if 3 consecutive exams yield a sight-playing accuracy equal or greater than 80%, we consider the learning goal achieved and the experiment is terminated. We first evaluate the difficulty of the pieces based on note density and pitch intervals [15] and make the curriculum alternates between easy and difficult songs. Such design avoids a "too early termination" caused by three consecutive easy exams. It also avoids a poor learning experience caused by three consecutive difficult ones.

### 4.2 Music Pieces to Learn
To better serve the task of elementary flute sight-playing using the Rainbow Score system, all 16 folk songs are modified to match three standards. First, pieces are rearranged to C major diatonic scale and within one octave, so that the fingering position can fully decide the pitch without measuring the breathing. Second, no two adjacent notes are of the same pitch, so that the hardware has no need to measure tongue movement. Finally, most pieces are eight measures to fit in a whole page of the Rainbow Score interface.

Piece No. 1

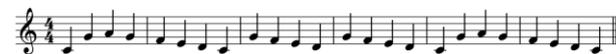

Piece No. 2

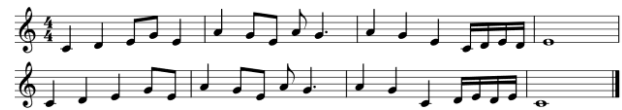

Piece No. 3

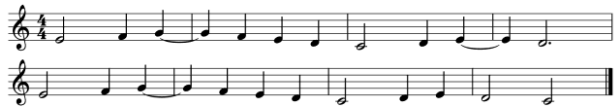

Piece No. 4

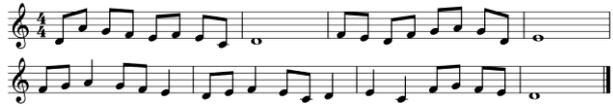

Piece No. 5

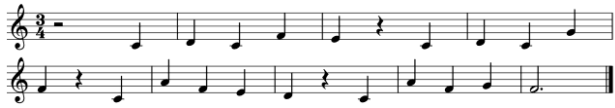

Piece No. 6

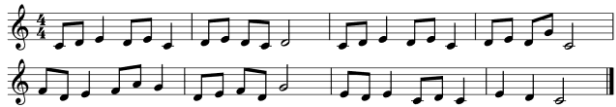

Piece No. 7

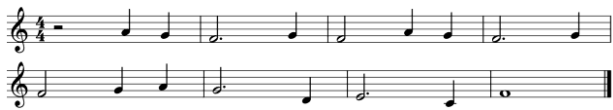

Piece No. 8

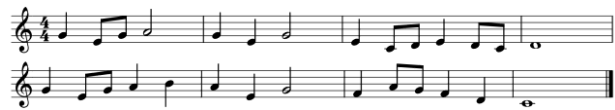

Piece No. 9

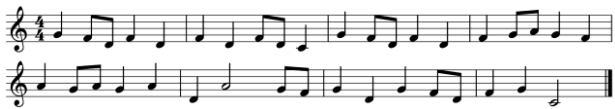

Piece No. 10

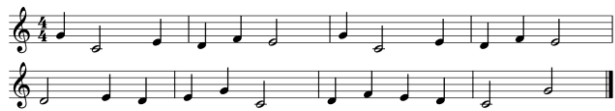

Piece No. 11

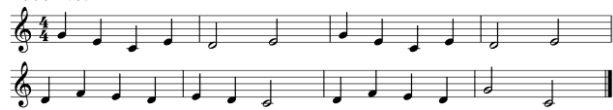

Piece No. 12

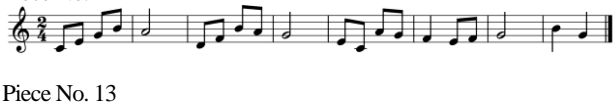

Piece No. 13

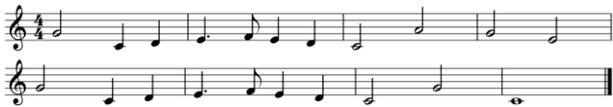

Piece No. 14

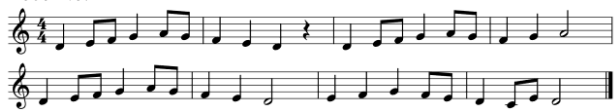

Piece No. 15

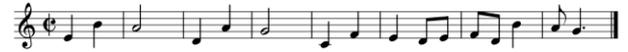

Piece No. 16

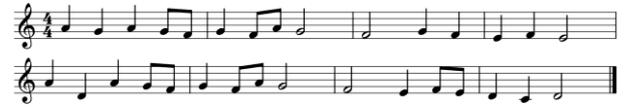

### 4.3 Participants

18 subjects (12 males and 6 females, age range 18-25, age median 20) participated in the study. They are randomly assigned to two groups: *interactive* vs. *static* (the control group). The former has access to learning modes A & B, while the latter only has access to learning modes C & D. No subject has prior experience on flute performance or sight-playing.

### 4.4 Task and Procedure

For each song, the subject goes through 4 steps: pre-exam, listen to the ground truth, practice, and randomized pitch exam. The subject is allowed to quit the experiment at any time.

**Pre-exam**: We test the subject to play the piece in learning mode C without any prior practice. If the performance score is equal or greater than 80%, the subject can choose to skip the song. The score of performance is computed as the number of correctly played notes divided by the total number of notes. A note is correctly played if the subject plays it correctly for $\geq$ 70% of its duration.

**Listen to the Ground Truth**: We play the ground truth (the correct performance of the piece).

**Practice**: The subject practices the piece using the modes available to him/her. The subject can request to listen to the ground truth again at any time. The practice time is limited to $\leq$ 15 minutes.

**Randomized Pitch Exam**: We modify the piece by randomizing the pitch of all notes while keeping the rhythm unchanged, and use the modified piece as another sight-playing exam. The purpose of the randomized pitch exam is to enforce that the notes can only be retrieved from the visual information but not the musical context. Both the score of the pre-exam and the score of the randomized pitch exam are recorded.

### 4.5 Results

We analyze the overall performances of the two groups in Section 4.5.1, study how the performance evolves over the training process in Section 4.5.2, and further consider individual difference in music talents in Section 4.5.3. We collected 16 valid results among the 18 invited subjects, 8 for each group. (Two quit too early so their learning curves are not informative.)

#### 4.5.1 Comparison of Average Learning Efficiencies

We define *learning efficiency* as the reciprocal of the total number of exams a subject takes before he/she passes the training. Since there are 16 songs in the curriculum, we have 32 total exams (half pre-exams and half randomized pitch exams). Therefore, the minimum score is $1/32$ and the maximum score is 1.

| Table 2. A comparison on average learning efficiency (1/Number of exams a subject took before success) |||
|---|---|---|
| Mode A & B | Mode C & D (baseline) | Improvements |
| **0.059** | 0.045 | 31.1% |

Table 2 shows that the average learning efficiency using Mode A & B is **31.1%** higher than that using Mode C & D, with $p = 0.084$ in independent t-test. This improvement shows that the interactive visual feedback accelerates the learning process.

### 4.5.2 Comparison of Learning Curves

Figure 5 shows how the average exam score (over 8 subjects) of each group evolves during the learning process. For both groups, we see a general ascending trend **and Mode A & B yields a better performance than Mode C & D** for almost all exams.

Note that both curves go up and down because curriculum alternates between easy and difficult songs. When a subject succeeds and leaves the experiment, the remaining part of the score is filled with 100%; if the subject quits, the last valid score is copied to fill the remaining scores.

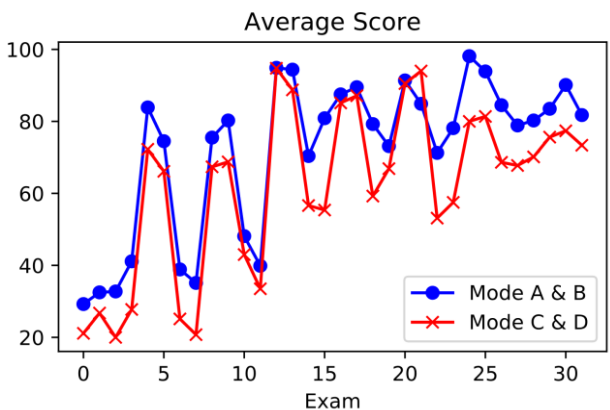

**Figure 5. A comparison on average exam scores throughout the learning process.**

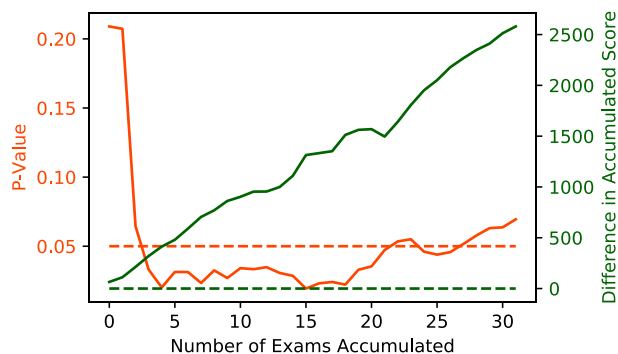

**Figure 6. Accumulated scores difference with p-value.**

Figure 6 further demonstrates the accumulated difference (green curve) between the two curves in Figure 5 and the corresponding p-value (orange curve) of t-test. It is interesting to see that, in general, the difference between the two groups is **more significant in the first half of learning**, which indicates that it is more critical to apply the interaction feature in the *early period* of learning.

### 4.5.3 Individual Differences

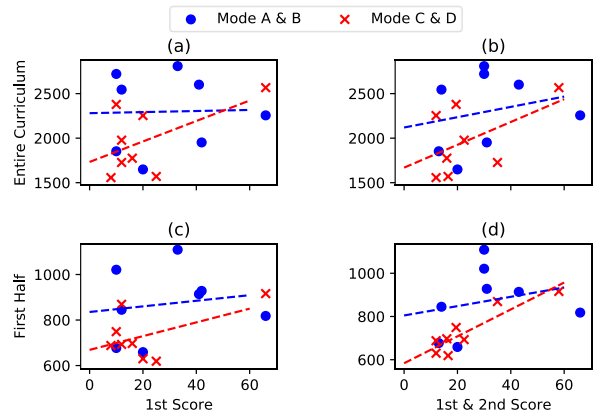

**Figure 7. The relationship between initial and overall performance.**

One may argue that the huge difference between the two groups may be caused by individual differences in the music talents, and subjects using Mode A & B are by chance more musical.

To rule out this possibility, we further examine the relationship between the initial talent and the overall performance and show it in Figure 7. In each subfigure, one dot represents a subject, with its x-coordinate being the subject's initial talent and its y-coordinate being the subject's overall performance. For the first row [(a) and (b)], the overall performance is measured by the accumulated score of the whole curriculum, while for the second row [(c) and (d)], overall performance is based on only the first half of the curriculum. Similarly, the graphs are divided into two columns based on what we use to represent the subject's initial talent: (a) and (c) use the $1^{st}$ exam score, while (b) and (d) use the sum of the $1^{st}$ and the $2^{nd}$ exam scores.

We see that for all subfigures, blue dots are on average above the red ones, especially for lower initial scores. This result indicates that interactive visual feedback is especially helpful for the less talented people, making the learning of music less dependent on personal talent. Moreover, the two groups of dots are further apart from each other on the second row, which means interactive features are more effective in the early period of learning.

### 4.5.4 Interview

Here, we report several interview questions and some representative answers to help gain a deeper understanding of the interactive feature.

**Q1. What do you think the learning process would be like if the interactive feature was unavailable to you? (Only asked to the group using Mode A & B)**

- *It* would *be more difficult, since I couldn't adjust myself.*

- *It would be slower. I wouldn't be able to tell if I played correctly.*

- *I would learn slower, but maybe I would gain a more thorough understanding.*

**Q2. What do you think the learning process would be like if the interactive feature was available to you? (Only asked to the group using Mode C & D. They try Mode A & B after the experiment)**

*- It would be faster. I would be able to tell apart different notes sooner.*

*- It would help me correct my mistakes. But I may be more nervous to sight-play.*

*- It would help my rhythm. I would realize I played every note a little bit late. However, the white masks are a little messy. Too much stuff on the screen can be distracting.*

**Q3. Do you have any other comments?**

*- In Mode A, I can hardly keep up with the music, let alone to look at the white masks. In Mode B, however, there is more time for me to read the white masks.*

*- I cannot hear my mistakes. I always have to spot them. Even when the interactive feature is off, I spot my mistakes by seeing that the color of the note does not match with the note I'm playing.*

*- In the beginning, I translated the visual directly to my finger motions. Later, the translation acquired an intermediate step: the abstract musical note. From this point I could identify my mistakes using my ears.*

*- The white masks help because it parallels with the finger positions. I immediately identify my mistakes when I make them. I use the white masks to know which direction to incrementally correct my mistakes. I can hear my mistakes.*

*- I use height information to identify note C - F, and color information to identify note G - B.*

*- I use color to remember that pink notes are A. The color helps on a subconscious level.*

## 5. DISCUSSION

The remarks from the users confirm our hypothesis of why the interactive feature is effective. With the real-time feedback, the player learns the mapping between motion and visuals not only when he/she plays correctly, but also when he/she makes mistakes. This augments the learning material beyond the original piece and prevents the learner from ever losing track of what note he/she is playing. Moreover, the learner and the tutoring system form a loop where the learner tries to translate visual notations to motions and the system translates motions to visuals back for the learner to improve such translation ability. Furthermore, graphics are intrinsically less abstract than a time series of performance motions or pitches, which makes the visual channel suitable for displaying performance mistakes.

## 6. CONCLUSION AND FUTURE WORK

We have contributed Interactive Rainbow Score, an interactive visual interface to assist elementary sight-playing training. We found that the interactive visual feedback on average boosts the learning efficiency by 31.1%. Based on the observation of the learning curves, we conclude that the benefit of interactive visual feature is especially significant in the early stage of learning.

Above all, the new interface sheds some light on learning general musicality using interactive systems.

In the future, we will continue to investigate multimodal interactive strategies for music learning. Firstly, we plan to integrate the visual feedback feature with haptic-based systems. More generally, we would like to build a theoretical model that explains the interactions between various modal inputs and outputs of a learner.

## 7. ACKNOWLEDGMENT

We want to thank Larry Wasserman, Xingdong Yang, and Ziyu Wang for their feedback and recommendations.

## 8. REFERENCES

[1] Rock Band. www.rockband4.com.
[2] Dan Ellis. Chroma Feature Analysis and Synthesis. labrosa.ee.columbia.edu/matlab/chroma-ansyn.
[3] Katsuya Fujii, et al. ACM 2015. "MoveMe: 3D haptic support for a musical instrument." In *Proceedings of the 12th International Conference on Advances in Computer Entertainment Technology.*
[4] Graham Grindlay. 2008. Haptic guidance benefits musical motor learning. In *HAPTICS '08: Proceedings of the 2008 Symposium on Haptic Interfaces for Virtual Environment and Teleoperator Systems*, pages 397–404, Washington, DC, USA, 2008. IEEE Computer Society.
[5] Guitar Hero. www.guitarhero.com/game.
[6] Kevin Huang, et al. ACM, 2010. Mobile music touch: mobile tactile stimulation for passive learning. In *Proceedings of the SIGCHI Conference on Human Factors in Computing Systems.*
[7] K. Itoh, H. Sakata, I.L. Kwee, et al. Musical pitch classes have rainbow hues in pitch class-color synesthesia. Sci Rep 7, 17781 (2017).
[8] Jeanine Mae Jacobson. 2006. *Professional Piano Teaching: A Comprehensive Piano Pedagogy Textbook for Teaching Elementary-level Students.* Vol. 1. Alfred Music Publishing.
[9] O'Malley, M. K., & Gupta, A. 2008. *Haptic Interfaces.* Pages 25–74, Morgan-Kaufman Publisher.
[10] The One Smart Piano. www.smartpiano.com.
[11] Cyber Glove Systems. www.cyberglovesystems.com.
[12] Taiko no Tatsujin. www.nintendo.com/games/detail/taiko-no-tatsujin-drum-n-fun-switch.
[13] Larry Wang. Piano Roll. web.mit.edu/larryw/www/pianoroll.
[14] Gus Xia, Jacobsen, C., Chen, Q., Yang, X. and Dannenberg, R. 2018. ShIFT: A Semi-haptic Interface for Flute Tutoring. In *The 18th International Conference on New Interfaces for Musical Expression.*
[15] Yian Zhang, Yinmiao Li, Daniel Chin, and Gus Xia. 2019. Adaptive Multimodal Music Learning via Interactive Haptic Instrument. *Proceedings of the International Conference on New Interfaces for Musical Expression*, UFRGS, pp. 140–145.